\documentclass[aps,twocolumn,epsfig,graphics,floatfix,mathbbm,prl]{revtex4}

\usepackage{amsmath,amsfonts,amssymb,graphics,graphicx,epsfig,color,times,bbm}

\begin{document}
\bibliographystyle{apsrev}

\title{Creating single time-bin entangled photon pairs}

\author{Christoph Simon}
\email{christoph.simon@ujf-grenoble.fr}
\author{Jean-Philippe Poizat}

\affiliation{
 Laboratoire de Spectrom\'{e}trie
Physique, CNRS et Universit\'{e} Joseph\ Fourier - Grenoble, 140
rue de la Physique, BP 87, 38402 St.\ Martin d'H\`{e}res, France}

\date{\today}

\begin{abstract}
When a single emitter is excited by two phase-coherent pulses with
a time delay, each of the pulses can lead to the emission of a
photon pair, thus creating a ``time-bin entangled'' state. Double
pair emission can be avoided by initially preparing the emitter in
a metastable state. We show how photons from separate emissions
can be made indistinguishable, permitting their use for
multi-photon interference. Possible realizations are discussed.
The method might also allow the direct creation of $n$-photon
entangled states ($n>2$).

\end{abstract}


\maketitle

The development of convenient sources of entangled photons is an
important task in quantum information \cite{qibook}. Entangled
photons have been used to realize fundamental quantum information
procedures such as quantum key distribution
\cite{qkdvienna,qkdgeneva,qkdlanl}, quantum teleportation
\cite{tpvienna,tprome,tpgeneva} and entanglement swapping
\cite{entswap}. The latter task is an essential element of quantum
repeaters \cite{repeaters}, which would allow the distribution of
entanglement over very long distances. Recently entangled photons
were also used to implement simple quantum logic gates
\cite{photongates}.

The standard source of entangled photons at the moment is
parametric down-conversion \cite{qibook}, which is based on the
conversion of pump photons into pairs of photons inside a
non-linear optical crystal. An important drawback of
down-conversion sources is the fact that they cannot be made to
produce exactly one pair of photons. They always generate a
statistical distribution of pairs. If the probability to create a
single pair with a given pump pulse is $p$, then there is a
probability of order $p^2$ to create two or more pairs with the
same pump pulse. This feature of down-conversion sources leads to
limitations on their performance for various quantum information
procedures, such as teleportation \cite{teleportproblems}, quantum
cryptography \cite{qkdproblems} and entanglement purification
\cite{pdcpurification}. It would thus be very desirable to have a
convenient source of individual pairs of entangled photons, where
one can be sure that no more than one pair is emitted.

A natural approach towards realizing such sources is to use
photonic cascades from atoms or semiconductor quantum dots. Atomic
cascades were used to produce polarization-entangled photons in
the first tests of Bell inequalities \cite{aspect}. Quantum dot
sources are attractive because they are compact and can be fairly
easily integrated into semiconductor microcavity structures to
enhance the probability for emission of the photons into a
well-defined mode. These features have recently been demonstrated
for single-photon sources \cite{yamamoto}. There is a recent
proposal for a quantum dot source of single pairs of polarization
entangled photons \cite{benson}, based on the biexciton-exciton
cascade. However, the generation of polarization entanglement with
this source requires the two intermediate exciton states with
different spin to be exactly degenerate, which is not the case for
currently available quantum dots, due to their lack of exact
rotational symmetry around the direction of growth. In
consequence, current quantum dots can emit
polarization-correlated, but not polarization-entangled, photons
\cite{excitonsplitting}.

Here we propose to create single pairs of {\it time-bin} entangled
photons from single emitters such as atoms or quantum dots.
Time-bin entanglement was first introduced for down-conversion
sources in Ref. \cite{brendel}, based on the principle of Ref.
\cite{franson}. It requires a source that can generate a pair of
photons at two different well-defined times. This creation has to
happen coherently, such that no information about the time of
emission of the photon pair is stored anywhere in the emitting
system or in the environment. Under this condition, the generated
state is a superposition of both photons having been emitted at
the earlier or later time, of the schematic form
\begin{equation}
|\mbox{early}\rangle |\mbox{early}\rangle + |\mbox{late}\rangle
|\mbox{late}\rangle, \label{tb1}
\end{equation}
which is clearly an entangled state. Methods for detecting and
using this form of entanglement were first described in Ref.
\cite{brendel}.
In recent years time-bin entanglement has been successfully used
to implement various quantum information tasks
\cite{tpgeneva,qkdgeneva,gisinpapers}. It is particularly well
suited for long-distance transmission in optical fibers because it
is insensitive to polarization fluctuations.

\begin{figure}
\resizebox{0.6\columnwidth}{!}{\includegraphics*{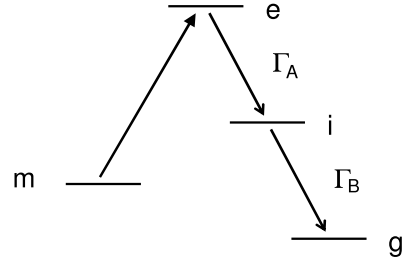}}
\caption{Level scheme for the proposed source of single time-bin
entangled photon pairs.
} \label{levels}
\end{figure}

Our proposed source emits at most one time-bin entangled photon
pair. This property is achieved by using a single two-photon
emitter, which is initially prepared in a metastable state to
eliminate the possibility of creating two pairs, as we will now
explain. Fig. 1 shows the required level scheme. There are three
levels $e$, $i$ and $g$ in a cascade configuration, plus a
metastable excited state $m$. If the system is prepared in the
excited state $e$, it quickly emits two photons, one each from the
transitions $e\rightarrow i$ and $i \rightarrow g$. For
entanglement generation, the system is first promoted to the
metastable state $m$. Then the system is made to interact with two
pump pulses that are in resonance with the transition from $m$ to
$e$. The two pulses have a fixed time delay and relative phase.
The first pulse excites the system to $e$ with a probability
$p_1$. If the system is excited, it quickly emits a pair of
photons and goes to the state $g$. In this case, the second pump
pulse has no effect, because it is far out of resonance. Starting
from the metastable state $m$ instead of the ground state $g$ thus
ensures that at most one pair of photons is emitted by the source.
The second pulse can excite the system from $m$ to $e$ with
probability $p_2$. Thus, if the system did not emit a pair of
photons after the first pulse, it can emit a pair after the second
pulse. Provided that the pumping is on resonance and that there is
no coupling to the environment, no information about the time of
emission of the photons remains in the system. The process thus
creates a pair of photons in a superposition of having been
emitted at two different points in time. Schematically, the
described procedure creates an entangled two-photon state
\begin{equation} \sqrt{p_1} |\mbox{early}\rangle |\mbox{early}\rangle + e^{i\phi_P} \sqrt{(1-p_1)p_2}|\mbox{late}\rangle |\mbox{late}\rangle, \label{tbp}
\end{equation} where $\phi_P$ is the relative phase between the two pump pulses.
The squared norm of the state (\ref{tbp}) describes the overall
probability to produce a photon pair. For $p_1=1/2$ and $p_2=1$
the source produces exactly one maximally entangled pair.

For a real system, the schematic wave functions of Eqs.
(\ref{tb1},\ref{tbp}) are not completely accurate. In particular,
the components of the state referring to an individual cascade,
which were denoted by $|\mbox{early}\rangle |\mbox{early}\rangle$
and $|\mbox{late}\rangle |\mbox{late}\rangle$ in the idealized
discussion above, are a priori not of the simple product form
suggested by the notation, but are themselves entangled states.
Below we will explain why this is a potential problem for
multi-photon interference experiments such as entanglement
swapping, and show how it can be solved. First we describe the
nature of the unwanted entanglement. It arises because there is a
time ordering of the two transitions in a cascade - the $e
\rightarrow i$ transition has to occur before the $i \rightarrow
g$ transition. This leads to a temporal correlation between the
two photons. The two-photon wave function referring to a single
cascade is \cite{scully,huang}
\begin{eqnarray} \psi(t_A,t_B)=
2 \sqrt{\Gamma_A \Gamma_B}  e^{-\Gamma_A t_A} \theta(t_A)
e^{-\Gamma_B (t_B-t_A)} \theta(t_B-t_A). \label{cascade}
\end{eqnarray}
Here the indices $A,B$ refer to the first and second photon from
the cascade, $t_{A,B}$ are the emission times (relative to the
excitation at $t=0$), and $\Gamma_{A,B}$ the decay rates for the
two transitions (cf. fig. 1). This wave function is entangled due
to the presence of the factor $\theta(t_B-t_A)$ that describes the
temporal ordering of the emission events in the cascade. (The
function $\theta(x)$ is equal to 1 for $x\geq 0$ and equal to 0
for $x<0$.)

This entanglement is not problematic for applications where only a
single pair is used at any given time, such as quantum key
distribution. However, it becomes an issue for entanglement
swapping and related applications. Entanglement swapping between a
pair of time-bin entangled photons labelled $A_1$ and $B_1$ and
another time-bin entangled pair labelled $A_2$ and $B_2$ proceeds
by detecting the photons $B_1$ and $B_2$ at an intermediate
location in such a way that any information about their origin is
erased, see fig. 4 of Ref. \cite{brendel} and the related
discussion. Note that photons $B_1$ and $B_2$ have to come from
the same stage of their respective cascades, otherwise they could
be distinguished by their frequencies. Also, for single-pair
sources as the one proposed here, in contrast to down-conversion
sources \cite{teleportproblems}, there is a vanishing probability
that both photons detected at the intermediate location were
emitted by the same source. The erasure of information about the
origin of the two photons $B_1$ and $B_2$ can only be perfect if
the arrival times of photons $A_1$ and $A_2$ contain no
information about which of the two photons detected at the
intermediate location was $B_1$ and which was $B_2$. This requires
that there is no temporal correlation between the photons in each
pair ($A_1 B_1$ and $A_2 B_2$) apart from the time-bin
entanglement, which means that the wave function (\ref{cascade})
has to have product form.

The problem can be analyzed in terms of wave function overlap. For
the entanglement swapping to work perfectly, the two photons $B_1$
and $B_2$ have to have perfect overlap. If photon $B_1$ is
entangled with $A_1$ and photon $B_2$ with $A_2$ (in addition to
the time-bin entanglement), then their quantum states given by
tracing over the $A$ photon in Eq. (\ref{cascade}) are not pure,
but mixed: $\rho_B=Tr_A |\psi_{AB}\rangle \langle \psi_{AB}|$.
Here we assume that the two independent sources of entangled pairs
are otherwise identical, such that they are described by the same
two-photon wave function. The average overlap of two systems
described by identical mixed states $\rho_B=\sum_i p_i
|\chi_i\rangle \langle \chi_i|$ (with $\chi_i$ the eigenstates of
$\rho_B$) is equal to $\sum_i p_i^2$, since the overlap is one if
the systems are both in the same eigenstate $\chi_i$ (which
happens with probability $p_i^2$), and zero otherwise. The latter
expression is equal to $\mbox{Tr} \rho^2_B$, whose departure from
unity therefore gives the order of magnitude of the expected error
in the entanglement swapping. A straightforward calculation using
Eq. (\ref{cascade}) gives
\begin{equation}
1-\mbox{Tr} \rho_{B}^2=\frac{\Gamma_B}{\Gamma_A+\Gamma_B}.
\label{error}
\end{equation}
From Eq. (\ref{error}) one sees that the error can be made small
by making the decay rate for the first transition in the cascade,
$\Gamma_A$, much bigger than that for the second transition,
$\Gamma_B$. (It is easy to see intuitively that the correlation
between the photons is very strong in the opposite case of
$\Gamma_B \gg \Gamma_A$ because then the uncertainty of their time
difference is much smaller than the uncertainty of the emission of
each photon individually.)
If $\Gamma_A$
is not much bigger than $\Gamma_B$ in a given system, the decay
rates can be influenced through external cavities, by having a
more significant Purcell effect for the first transition.

The problem of unwanted temporal entanglement also exists for
down-conversion sources, although the entanglement has a slightly
different character. For these sources, which are very broad in
energy since the emission process is extremely fast (on a fs
timescale), the wave function $\psi(t_A,t_B)$ can be made to have
essentially product form by frequency filtering
\cite{tpvienna,tpgeneva}. This approach does not work for single
emitters as discussed here because they are already very narrow in
frequency. However, the fact that the emission can be
comparatively slow for single emitters (on a ns timescale) may
allow an approach based on time-resolved detection, which could be
used in some situations where tuning the decay times is not
feasible, or to achieve even better indistinguishability. This
second approach is based on noticing that the wave function
(\ref{cascade}) is of product form apart from the factor
$\theta(t_A-t_B)$. Therefore, if one detects photons of type $A$
only in an interval $[0,T_1]$, and photons of type $B$ only in an
interval $[T_2,T_2+\Delta T]$, where $T_2 > T_1$, then the wave
function is projected onto a perfect product state in the space
defined by the two intervals. The timing of one photon then
carries no information whatsoever about its partner. This
remarkable property of the wave function (\ref{cascade}) is a
consequence of the exponential behavior of the photon emission.

We will now discuss the perspectives for an experimental
realization of our proposal. As mentioned in the introduction,
atomic cascades were used in the first Bell experiments
\cite{aspect}. Recently a single-photon source that generates
photons in a well-defined radiation mode was realized with atoms
in a high-finesse cavity \cite{atsphsources}. Metastable states as
required in the level scheme of fig. 1 certainly exist in atoms,
for example the $D_{3/2}$ and $D_{5/2}$ states in the
$^{40}$Ca$^+$ ions used in recent quantum information processing
experiments \cite{blatt}. These ion traps have also been
integrated with high-finesse cavities \cite{eschner}. Compared to
quantum dots, the decay times in atoms are typically longer. For
decay times of the order of ns and longer, the time-resolved
detection described above becomes realistic. The most commonly
used avalanche photo diodes have a time resolution of the order of
hundreds of ps. However, new single-photon detectors based on
superconductors can achieve a time resolution as fine as 18 ps
\cite{supdet}. We believe that realizing the proposed protocol
with atoms is an interesting possibility that deserves further
investigation.

Here we will focus on implementing the proposal with quantum dots.
As mentioned in the introduction, a single-mode single-photon
source was recently realized based on a quantum dot embedded in a
micro-pillar shaped semiconductor micro-cavity \cite{yamamoto}.
The requirement that no information about the emission process may
leak into the environment (which is essential for creating the
time-bin entanglement, cf. above), is similar to the requirements
for the emission of individual photons into a single temporal
mode, which was demonstrated using the above-mentioned type of
source in Ref. \cite{yamnat}. We have also mentioned before the
recent proposal to create polarization entangled photon pairs via
exciting the bi-exciton state in a quantum dot \cite{benson}. The
resulting two-photon cascade via the single-exciton states is a
natural candidate for our proposal as well. For the present
source, one should select definite polarizations for the two
photons, thus reducing the cascade to a single intermediate state.
The correlation in polarization between the two emitted photons
was demonstrated in Ref. \cite{excitonsplitting}.

The metastable state $m$ can be realized by using a dark exciton
state, which is connected to the quantum dot ground state by an
optically forbidden transition ($\Delta J_z=\pm 2$). The lifetimes
of such dark excitons are orders of magnitude larger than those of
bright (optically allowed) excitons \cite{darkex}. Preparing the
system in the metastable state can simply be done by excitation to
the conduction band with subsequent relaxation. The system will
relax to a dark exciton state with a probability of order 1/2.
Exciting the system from the dark exciton state to the bi-exciton
state requires driving a transition that is optically forbidden.
However, it should be possible with realistic light intensities,
since in real quantum dots there is always some coupling of dark
and bright excitons due to valence band mixing. For example,
radiation from the dark exciton was recently observed in II-VI
quantum dots, showing that the transition is not strictly
forbidden \cite{besombesprb}. In the same experiment, the ratio of
the lifetimes of the dark and bright exciton states was determined
to be of order 100 \cite{besombesphd}. Similar results are
expected for III-V quantum dots, where discrimination of the
radiation from the dark exciton is more difficult because of a
smaller splitting between the dark and bright exciton levels.

Typical lifetimes for the bi-exciton to exciton and the exciton to
ground state transitions in III-V quantum dots (such as InAs) are
of order 0.6 ns and 1.4 ns respectively \cite{moreau}. The decay
rate $\Gamma_A$ is thus larger than $\Gamma_B$ already without a
cavity. In order to further reduce the unwanted temporal
entanglement, this ratio could be significantly enhanced by
embedding the quantum dot in a micro-cavity as in Ref.
\cite{yamamoto}. For example, for the above values of the decay
rates, a Purcell factor of 20 for the first transition in
combination with a Purcell factor of 2 for the second transition
would already reduce the error due to the temporal correlations
given by Eq. (\ref{error}) to the level of 5 percent. A Purcell
factor of order 6 was reported in Ref. \cite{yamamoto}. It should
be possible to achieve such a combination of Purcell factors with
a single micro-cavity, since the frequency difference between the
two transitions (which is due to the exchange interaction between
the excitons in the bi-exciton state) is of the same order of
magnitude as a typical micro-cavity linewidth. For example, the
splitting between the exciton and bi-exciton lines in Ref.
\cite{moreau} is of order 3 meV (or 2 nm), whereas the cavity
linewidth in Ref. \cite{solomon} is of order 2 meV. For an
appropriately chosen quantum dot and micro-cavity, one should thus
be able to bring the bi-exciton to exciton line into exact
resonance with the cavity (by temperature tuning) to maximize the
Purcell effect, while still achieving a smaller Purcell effect for
the exciton to ground state transition.

Demonstration experiments with a quantum dot source would proceed
in analogy to the experiments and setups described in Ref.
\cite{brendel}, which were inspired by the proposal of Ref.
\cite{franson}. For example, the setup for verifying the presence
of time-bin entanglement contains three interferometers with
identical path length difference and adjustable phases between the
two paths. The first one is placed in the pump beam to create the
two pump pulses from the output pulse of a mode-locked laser.
There is also an interferometer at each observer station, $A$ and
$B$. After their emission, the photons can be split by a suitable
wavelength-sensitive element, and photon $A$($B$) is directed to
observer $A$($B$). Interference occurs between the possibility
that the photons were created by the first pump pulse, and then
both took the longer path in the interferometers at $A$ and $B$,
and the possibility that they were created by the second pump
pulse and then both took the shorter path in the interferometers
at $A$ and $B$. As a consequence the coincidence probabilities
between interferometer outputs at $A$ and $B$ vary sinusoidally
with the combined phase $\phi_P-\phi_A-\phi_B$, where $\phi_P$ is
the phase between the paths of the pump interferometer and
$\phi_{A(B)}$ is the phase of the interferometer in $A$($B$). Ref.
\cite{brendel} describes how time-bin entanglement can be used for
quantum key distribution, entanglement swapping and other
multi-photon interference experiments. Several of these proposals
were realized for down-conversion sources in Refs.
\cite{qkdgeneva,tpgeneva,gisinpapers}. The basic detection methods
described in these papers are equally applicable to our proposed
source.

Current experiments on light emission from quantum dots at the
single photon level typically use III-V (InAs) quantum dots with
exciton wavelengths around 900 nm
\cite{yamamoto,excitonsplitting,yamnat,moreau,solomon}. However,
InAs quantum dots have been shown to be capable of emitting around
1.3 $\mu$m \cite{ustinov}, and recently even close to 1.5 $\mu$m
\cite{dasilva}. There thus seems to be a real possibility of
realizing a source of single time-bin entangled photon pairs at
telecommunication wavelengths, which would be very valuable for
long-distance quantum communication.


Another very interesting perspective is the possibility to create
time-bin entanglement not only of photon pairs, but of larger
numbers of photons directly from a single emitter, e.g. a quantum
dot. This could be done by having the pump pulses in our scheme be
in resonance with the tri-exciton or even higher excitonic states
instead of the bi-exciton. This would create states of the form
\begin{equation}
|\mbox{early}\rangle_A |\mbox{early}\rangle_B
|\mbox{early}\rangle_C... + |\mbox{late}\rangle_A
|\mbox{late}\rangle_B |\mbox{late}\rangle_C..., \label{ntuplets}
\end{equation}
where again the ideal product form of the two terms could be
achieved by tuning of the decay rates or by time-resolved
detection. A three-photon cascade was demonstrated in Ref.
\cite{persson} based on two-photon coincidence measurements. The
demonstration and use of time-bin entanglement requires the
collection and detection of all the photons from the cascade.
Considering the external quantum efficiency of close to 40\%
reported for the single-photon source of Ref. \cite{yamamoto}, the
coincident detection of three or more photons does not seem at all
unrealistic with current single-photon detectors.
A source that produces entangled $n$-tuplets of photons directly
and efficiently would be of major interest for quantum information
processing and quantum communication.

We would like to thank L. Besombes, C. Couteau, and S. Hastings
for useful discussions and helpful comments. This work was
supported by the French Ministry for Research (ACI Jeune Chercheur
no. 2063).

\end{document}